# Pressure-regulated mechanochemistry at lithium metal–sulfide electrolyte interfaces

Kunik Jang[1], Jaehwan Choi[1], Jang Wook Choi[1,2,3], and Yousung Jung[1,4]*

[1] Department of Chemical and Biological Engineering (BK21 four), and Institute of Chemical Processes, Seoul National University, 1 Gwanak-ro, Gwanak-gu, Seoul 08826, Korea

[2] Institute for Battery Research Innovation (IBRI), Seoul National University, 1 Gwanak-ro, Gwanak-gu, Seoul 08826, Korea

[3] Seoul National University Energy Initiative (SNUEI), Seoul National University, 1 Gwanak-ro, Gwanak-gu, Seoul 08826, Korea

[4] Institute of Engineering Research, Seoul National University, 1 Gwanak-ro, Gwanak-gu, Seoul 08826, Korea

* Email: yousung.jung@snu.ac.kr

**Abstract**

Stack pressure is commonly treated as a means of maintaining physical contact in all-solid-state lithium-metal batteries, but it can also alter the chemistry of reactive solid–solid interfaces. Here, using pressure-aware, charge-resolved machine-learning molecular dynamics validated against DFT, we determine how pressure magnitude and loading geometry regulate interphase formation at Li||$Li_6PS_5Cl$ interfaces. The response is nonmonotonic: compression at 1 kbar accelerates $PS_4$ decomposition and $Li_2S$-like ordering, whereas 10–100 kbar compression restricts structural rearrangement and long-range crystallization. Charge-resolved dynamics further identify sulfur-centered, lithium-rich early-interphase environments associated with subsequent $Li_2S$-like ordering. Uniaxial loading accelerates interfacial reaction relative to isostatic loading at the same nominal pressure. Pressure also changes void closure and dead-lithium spreading in a defect-location-dependent manner. These results establish applied pressure as a mechanochemical process variable coupling interphase chemistry, ion transport and defect evolution, providing a mechanistic framework for interpreting pressure effects in sulfide solid-state batteries.

All-solid-state Li-metal batteries (ASSLMBs) have been studied as promising next-generation energy-storage systems because they can combine the high theoretical capacity of Li metal (3860 mAh $g^{-1}$) with nonflammable solid electrolytes and compact cell architectures[1, 2]. Among solid electrolytes, sulfide argyrodites such as $Li_6PS_5Cl$ (LPSC) are particularly attractive because of their high Li-ion conductivity[3, 4], mechanical softness, and facile processability into dense electrolyte layers[5, 6]. However, pairing Li metal with sulfide electrolytes creates a reactive solid-solid interface whose structure, chemistry, and mechanical integrity often govern the performance of sulfide-based ASSLMBs[7, 8].

The Li||LPSC interface is chemically unstable because metallic Li is a strong reductant and sulfide electrolytes possess reduction potentials above the $Li^+$/Li potential[9, 10]. Direct contact between Li metal and LPSC therefore drives Li oxidation and electrolyte reduction, producing a solid-electrolyte interphase (SEI) that can contain lithium sulfide, phosphide, chloride, and mixed decomposition products[11, 12]. An interphase that is chemically stable, electronically blocking, and sufficiently Li-ion conducting can suppress continued electrolyte reduction, while allowing Li-ion transport[13]. In contrast, continuous electrolyte decomposition produces an excessively thick SEI and resistive byproducts, consumes electrolyte and active Li, increases ion-transport barriers and impedance, and ultimately degrades interfacial stability[9, 11]. Thus, the central question is not simply whether an interphase forms, but how its local structure, composition, and transport properties evolve from the earliest stages of contact.

Traditionally, interphase formation in ASSLMBs has been examined using in situ and ex situ XRD, XPS, Raman spectroscopy, ToF-SIMS, and TEM[12, 14]. These techniques identify reaction products, chemical states, depth profiles, and morphology, but Li/sulfide contacts are buried, air-sensitive, and reactive immediately after contact. Experiments therefore provide spatially averaged signatures or post-reaction states, making it difficult to track atomic-scale interphase evolution and separate interphase-specific signals from those of adjacent Li metal and electrolyte[7, 12].

Pressure is applied during both the fabrication and operation of solid-state batteries to densify components and maintain particle–particle and electrode–electrolyte contact[15, 16, 17]. It is therefore commonly treated primarily as a mechanical parameter that improves physical contact and lowers interfacial resistance[16, 18]. Growing evidence, however, indicates that external pressure couples directly to the electrochemistry of solid-state cells: at chemically unstable lithium–sulfide interfaces, compression can alter local coordination, atomic mobility, charge redistribution and the structural rearrangements accompanying electrolyte reduction. How pressure magnitude and loading geometry affect these coupled chemical and morphological processes remains poorly understood.

Machine-learning interatomic potentials (MLIPs) now enable reactive molecular dynamics with near-DFT accuracy[19] at the length and time scales of interfacial chemistry[20, 21]. Large-scale simulations of this kind have revealed $PS_4$ decomposition and $Li_2S$-like interphase formation at lithium–sulfide interfaces[22, 23], and charge-equilibration extensions allow interfacial charge redistribution to be followed under electrochemical boundary conditions[24]. These charge-resolved schemes, however, have so far lacked the long-range electrostatic contribution to the stress tensor, restricting them to fixed-volume ensembles and leaving the response of reactive interfaces to applied pressure unresolved. Here we extend DFT-validated Deep Potential[25, 26] and charge-equilibration simulations[24] to pressure-controlled ensembles by incorporating the electrostatic contributions to the virial, and examine Li||LPSC interfaces across pressure magnitude, loading geometry and defect location.

## Results

### Validated simulations of pressure-dependent interface evolution

To simulate reactive Li||LPSC interfaces under pressure, we constructed DFT-trained DP[25, 26] and DP-QEq[24] models using AIMD and active-learning data[27] spanning Li metal, LPSC bulk and ordered, Li-vacancy-containing and S-rich interfaces (Fig. 1a, Supplementary Fig. 1, Supplementary Note 1 and Supplementary Table 1). The training set included NPT configurations from 1 to 10 kbar and active-learning exploration up to 150 kbar. DP was used for long-timescale interphase evolution, whereas DP-QEq provided atom-resolved charges under fixed-charge and constant-potential conditions. For pressure-controlled DP-QEq simulations, the QEq electrostatic contribution was included consistently in the energy, forces and virial.

Both models were evaluated against independent AIMD trajectories (Supplementary Note 2, Supplementary Figs. 2 and 3, and Supplementary Tables 2 and 3) and DFT-relabeled snapshots from 200 ps DP-driven NPT trajectories spanning 0 bar to 100 kbar. Across the production-validation set, DP yielded energy, force and virial RMSEs of 6.199 meV atom$^{-1}$, 84.21 meV Å$^{-1}$ and 6.663 meV atom$^{-1}$, respectively, while DP-QEq yielded 7.483 meV atom$^{-1}$, 90.22 meV Å$^{-1}$ and 7.681 meV atom$^{-1}$ (Fig. 1b,c and Supplementary Tables 4 and 5). DP-QEq re-evaluation also closely followed the DP potential-energy profiles (Supplementary Fig. 4). Pressure-specific errors remained comparable across the sampled range. DP and DP-QEq also reproduced AIMD structural and Li-transport trends, whereas the short-range DP component alone underestimated diffusivity, confirming the need for the electrostatic contribution (Supplementary Note 3 and Supplementary Figs. 5–11).

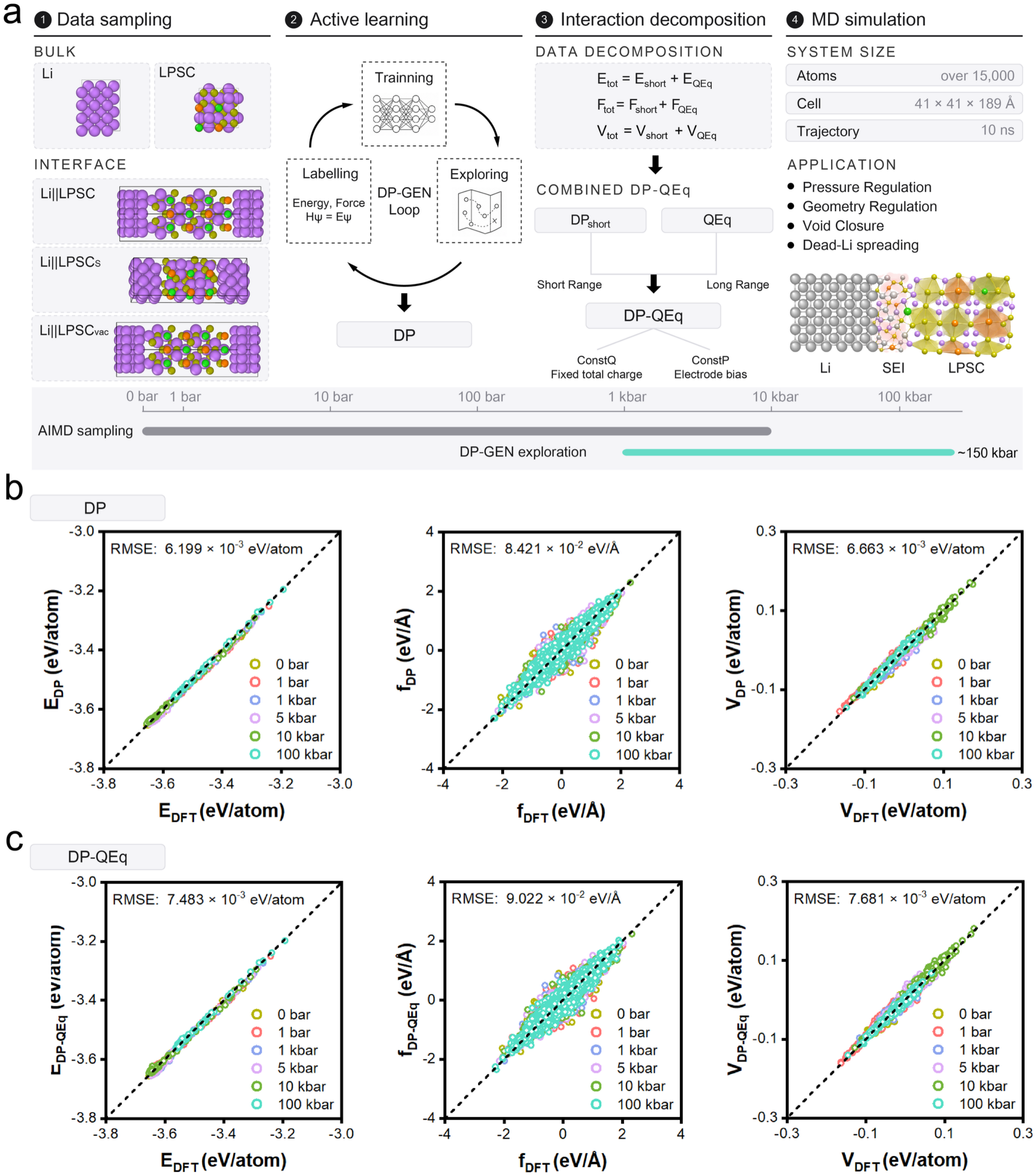

**Fig. 1 | Machine-learning interatomic-potential workflow and production-trajectory DFT validation for Li||LPSC interfaces under pressure.** (a) Overall workflow for constructing the pressure-aware DP and DP-QEq models: AIMD sampling of Li metal, LPSC bulk, and Li||LPSC interfacial configurations (Li||LPSC, Li||LPSC$_{vac}$, and Li||LPSC$_{s}$); active-learning enrichment by DP-GEN; decomposition of the energy, forces, and virial into DP$_{short}$ and QEq electrostatic contributions; and large-scale MD applications to pressure-, void-, and dead-Li-regulated interfacial phenomena. Energy, force, and virial parity plots comparing DFT labels with (b) DP and (c) DP-QEq ConstQ predictions for snapshots sampled from 200 ps DP-driven Li||LPSC NPT trajectories. The displayed RMSEs were calculated over all production-validation structures, while pressure-specific values are reported in Supplementary Tables 4 and 5.

### Nonmonotonic pressure regulation of interphase evolution

The low standard reduction potential of $Li^+$/Li (-3.04 V vs SHE) makes the Li‖LPSC interface intrinsically prone to electrolyte reduction. Thermodynamic reduction calculations[28, 29] show that LPSC is reduced against Li to form $Li_2S$, $Li_xP_y$, and LiCl (Supplementary Fig. 12), consistent with experimentally reported reduction products for sulfide electrolytes[9, 11]. To examine how these thermodynamically accessible reactions proceed under mechanical loading, we used the DP model for long-time simulations at 400 K from 1 bar to 100 kbar, with the elevated temperature used to access interfacial evolution within the simulated timescale. These conditions should not be interpreted as a continuous representation of practical stack operation. The 1 bar condition serves as a low-pressure reference, while 1 kbar (100 MPa) falls within the upper range of experimentally applied stack pressures[30]. The 10 kbar condition probes high-pressure densification and interfacial assembly[31], whereas 100 kbar defines an extreme-pressure mechanistic regime to isolate how strong compression affects framework decomposition and structural rearrangement. Moreover, although nominal stack pressures during operation are typically a few to tens of MPa, fractional contact at asperities, constriction sites and crack or dendrite tips concentrates such nominal loads into local stresses approaching the GPa-scale hardness of solid electrolytes. Local stress concentrations can make the 10 kbar condition mechanistically relevant to imperfect interfaces, whereas 100 kbar is used as an extreme-pressure limit rather than as a practical stack-pressure analogue.

In the 1 bar and 1 kbar trajectories, Li metal first reacts with the LPSC framework, $PS_4$ tetrahedra progressively decompose, and an amorphous interphase forms before nanocrystalline regions emerge (Supplementary Figs. 13 and 14). This amorphous-to-crystalline evolution agrees with experimental and MD observations[22, 23, 32] while showing that the extent of crystallization is clearly pressure-dependent. Crystalline interphase regions are apparent at 1 bar and 1 kbar, become spatially constrained at 10 kbar, and are not evident at 100 kbar within 10 ns (Fig. 2a).

The pressure dependence is nonmonotonic. After 10 ns, clear $Li_2S$-like reflections appear at 1 bar and 1 kbar, with the most intense and narrowest signal at 1 kbar, whereas the corresponding features remain broad and weak at 10 and 100 kbar and a larger fraction of the LPSC framework is retained (Fig. 2b,c and Supplementary Fig. 15). Three independently initialized trajectories at each pressure reproduce this trend, and a sharp $Li_2S$-like reflection remains absent when the 10 kbar trajectory is extended to 20 ns (Supplementary Fig. 16 and Supplementary Table 6). Thus, 1 kbar promotes interfacial conversion and $Li_2S$-like ordering relative to 1 bar, whereas stronger compression limits the structural rearrangement required for long-range crystallization.

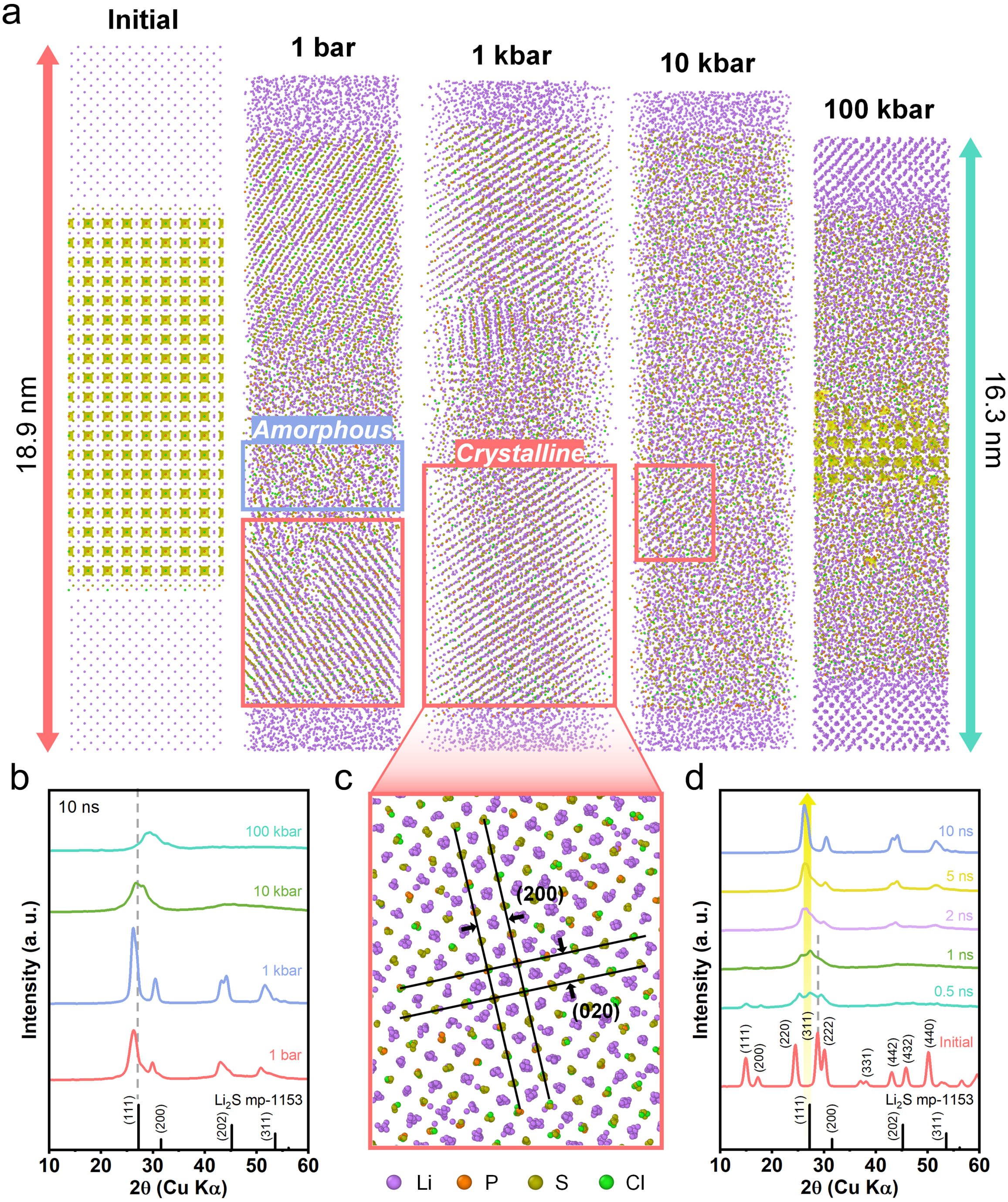

**Fig. 2 | Pressure-dependent long-time evolution of the Li|LPSC interphase using DP.** (a) Initial and 10 ns DP-NPT snapshots of the Li|LPSC interphase at 1 bar, 1 kbar, 10 kbar, and 100 kbar. Li, P, S, and Cl are shown in purple, orange, yellow, and green, respectively. (b) Simulated XRD patterns after 10 ns at different pressures, compared with the $Li_2S$ reference phase. (c) Enlarged view of the crystalline interphase formed at 1 kbar, highlighting a $Li_2S$-like framework in which P and Cl are incorporated into the anion sublattice. (d) Time-resolved XRD patterns at 1 kbar from the initial structure to 10 ns, showing the gradual emergence and growth of the $Li_2S$-like main reflection. The indexed reflections in the initial pattern correspond to the crystallographic planes of pristine LPSC.

Time-resolved XRD at 1 kbar shows a shift of the remaining LPSC reflections to higher diffraction angles, indicating lattice contraction[33], together with the gradual growth of a new $Li_2S$-like reflection (Fig. 2d). The reflection is shifted to lower diffraction angles relative to pristine $Li_2S$ (mp-1153), and the corresponding structure contains P- and Cl-bearing motifs within the anion framework (Fig. 2c).

Together with the longer Li–P than Li–S distances in the RDFs, these observations indicate an expanded, chemically modified $Li_2S$-like structure rather than phase-pure $Li_2S$ (Supplementary Fig. 17).

RDF analysis shows that P–S correlations weaken while Li–S and Li–P correlations grow at 1 kbar; these changes are less pronounced at 100 kbar, consistent with slower interfacial conversion (Supplementary Fig. 17). CN maps show decreasing P–S coordination and increasing S–Li, P–Li and Cl–Li coordination during interphase formation (Supplementary Figs. 18–21). S–Li coordination approaches eight (CN = 8 for $Li_2S$, mp-1153), characteristic of $Li_2S$-like environments, at all pressures. In contrast, near-eight P–Li and Cl–Li coordination is more frequent at 1 bar and 1 kbar, whereas at 10 and 100 kbar P–Li shifts towards Li-rich phosphide-like environments (CN = 11 for $Li_3P$, mp-736) and Cl–Li approaches LiCl-like coordination (CN = 6 for LiCl, mp-22905). These trends indicate a more mixed $Li_2S$-like interphase at lower pressure and increasingly distinct phosphide- and chloride-like environments under stronger compression. Pressure therefore changes not only lattice dimensions but also reaction extent, crystallinity and chemical partitioning within the reduced interphase.

**Charge redistribution accompanies the emergence of $Li_2S$-like ordering**

The XRD patterns obtained from long-time DP simulations in Fig. 2d show that a weak $Li_2S$-like reflection is already present at 0.5 ns, before the crystalline feature becomes dominant at 10 ns. This signal suggests early $Li_2S$-like ordering and motivates examination of the precursor to subsequent crystallization. Understanding how this precursor structure develops is therefore important for clarifying the origin of subsequent $Li_2S$-like crystallization. To examine this early stage with charge-resolved dynamics, we performed DP-QEq ConstQ simulations at 400 K over the first 400 ps. The QEq charge trends were validated against DFT-derived Hirshfeld and CM5 charge analyses for a smaller Li|LPSC structure (Supplementary Fig. 22). The absolute charge values differ because Hirshfeld, CM5, and QEq utilize different definitions of atomic charge, but the element-specific trends are consistent. This agreement supports the use of DP-QEq charges as qualitative descriptors of charge transfer and local redox evolution rather than as unique formal oxidation states.

Time-resolved XRD under the ConstQ condition shows that the $Li_2S$-like reflection, which appears between the initial LPSC (220) and (311) peaks in the long-time trajectory, begins to emerge within the first few hundred picoseconds (Fig. 3a). At 100 ps, the total XRD pattern does not yet contain a fully developed $Li_2S$-like peak, but region-specific XRD separates the signal of the early interphase from those of Li metal and the remaining LPSC bulk (Fig. 3c). The interphase region already shows a weak feature between the LPSC (220) and (311) reflections, indicating the formation of a structurally biased amorphous precursor rather than a fully crystalline phase. This early structural bias provides a plausible origin for subsequent $Li_2S$-like crystal growth.

Early $PS_4$ tetrahedra decomposition also shows a pressure-dependent acceleration and suppression (Supplementary Fig. 23). At 1 kbar and 10 kbar, the $PS_4$ destruction ratio increases faster than at 1 bar during the first 100 ps, whereas at 100 kbar the decomposition is strongly suppressed. Visualizations of the 1 bar and 100 kbar trajectories show that the amorphous interphase forms more rapidly and grows thicker under low pressure, while very high pressure preserves more of the LPSC framework during the same time window (Supplementary Figs. 24 and 25). The combined early-stage and long-time simulation results indicate that pressure affects initial framework decomposition and subsequent crystalline ordering differently. Compression at 10 kbar accelerates early $PS_4$ decomposition but permits only limited crystallization over longer times, whereas 100 kbar kinetically constrains the structural rearrangements required for continued framework breakdown.

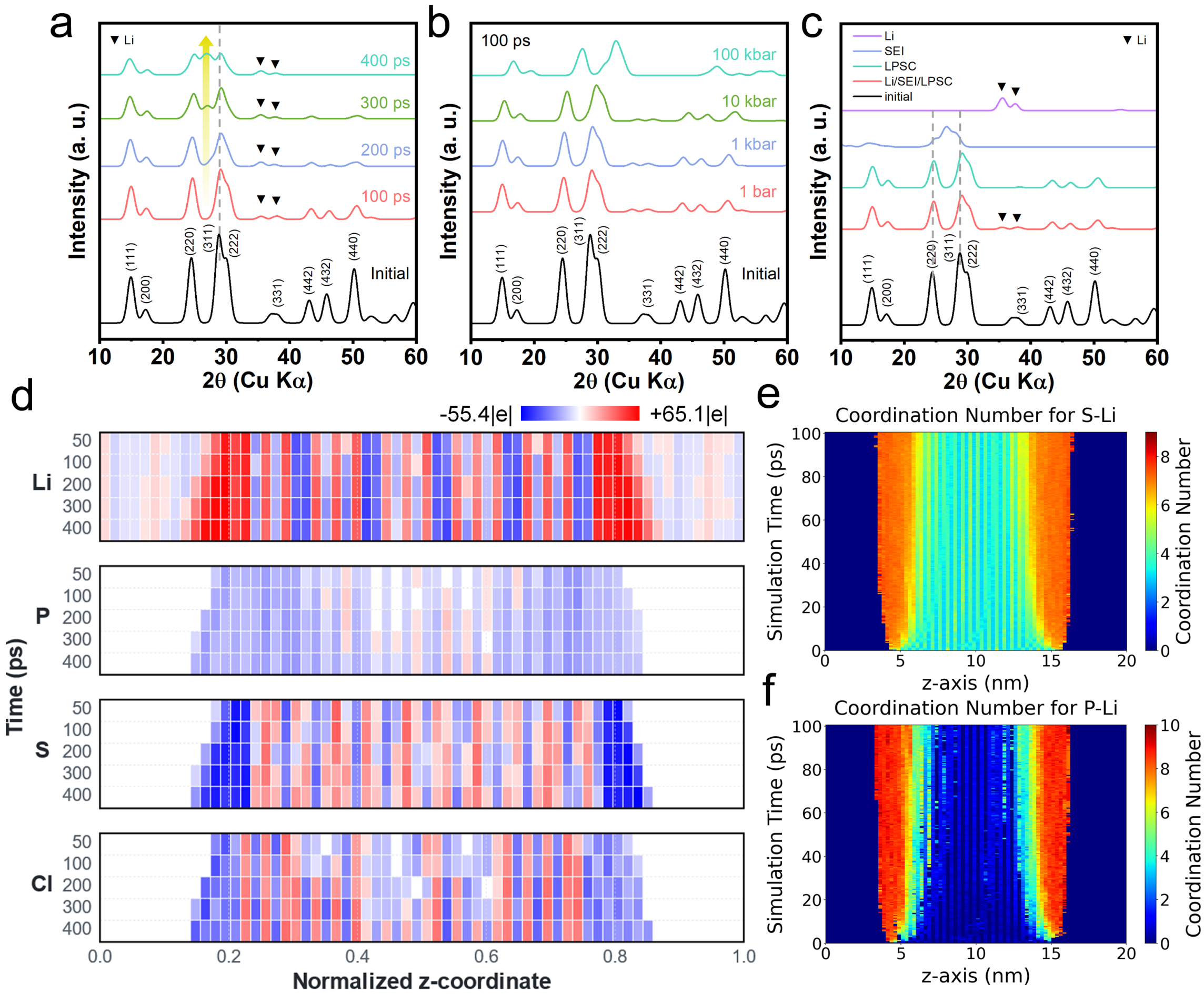

**Fig. 3 | Early-stage MD simulation of Li|LPSC interphase formation using DP-QEq ConstQ.** (a) Time-resolved XRD patterns of Li|LPSC from 0 to 400 ps at 1 bar. (b) Pressure-dependent XRD patterns at 100 ps from 1 bar to 100 kbar. (c) Region-resolved XRD patterns at 100 ps for Li metal, SEI, LPSC, and the total structure. The indexed reflections in the initial pattern correspond to the crystallographic planes of pristine LPSC. (d) Charge change of Li, P, S, and Cl relative to the initial structure during 400 ps at 1 kbar. (e,f) Time-resolved coordination number maps for (e) S-Li and (f) P-Li during the first 100 ps at 1 kbar.

Charge and coordination analyses identify the chemical nature of the early precursor. During the first 400 ps at 1 kbar, electron density is redistributed from Li metal toward the LPSC framework (Fig. 3d). Charge redistribution is element-selective, with S showing the strongest negative charge accumulation among LPSC elements and correlating spatially with positively charged interfacial Li. This preferential stabilization of S-centered, Li-rich local environments is consistent with the later emergence of $Li_2S$-like, rather than $Li_3P$- or LiCl-like, crystallinity in the long-time MD simulations. During the first 100 ps, S-Li, P-Li, and Cl-Li coordination increase as $PS_4$ units are disrupted and the interphase thickens (Figs. 3e,f and Supplementary Fig. 27). The dominant early S-Li coordination remains slightly below that of crystalline $Li_2S$ (CN = 8 for $Li_2S$, mp-1153), and P/Cl coordination is still broad, showing that the amorphous interphase is a mixed reduced phase rather than a set of already separated crystalline $Li_2S$, $Li_3P$, and LiCl domains. Pressure further modifies this early amorphous interphase by increasing local Li coordination around S, P, and Cl (Supplementary Figs. 26-29), consistent with densification that may influence the subsequent interfacial reaction pathway.

To determine whether pressure affects the constituent regions of Li||LPSC differently, we separately analyzed structural ordering in Li metal and Li-ion diffusivity in the amorphous interphase and crystalline LPSC. Structural classification shows that Li remains predominantly BCC up to 1 kbar, whereas the FCC fraction increases at higher pressures; isolated-region diffusivity calculations further show higher Li-ion mobility in the amorphous interphase than in crystalline LPSC under the analyzed high-temperature conditions. Increasing pressure promotes denser Li-metal packing while reducing Li diffusivity in both electrolyte-derived regions; this mobility decrease is consistent with restricted structural rearrangement and may help explain why 10 kbar accelerates early $PS_4$ decomposition while limiting subsequent long-range ordering (Supplementary Note 4 and Supplementary Figs. 30 and 31).

To assess the effect of electrochemical bias, DP-QEq ConstP simulations show that applied bias accelerates and spatially directs interfacial reduction, while the dominant coordination motifs remain similar to those under ConstQ over the simulated timescale (Supplementary Note 5 and Supplementary Figs. 32–36).

**Loading geometry redirects interfacial deformation and reaction**

Pressure-assisted densification is essential in solid-state electrolyte processing because it reduces internal voids and improves particle-particle and electrode-electrolyte contact[5, 16]. Recent experiments therefore examine both pressure magnitude and loading mode by comparing uniaxial and isostatic compression[34, 35]. However, such experiments mainly evaluate macroscopic cell performance, while the corresponding atomic-scale deformation, local contact evolution, and structural changes at interfaces remain difficult to probe directly. To isolate the effect of pressure-loading geometry, we compared 10 kbar isostatic and 10 kbar laterally constrained uniaxial DP-QEq ConstQ NPT simulations with isotropic and z-axis-only barostat coupling, respectively. In the uniaxial case, the lateral cell dimensions are held fixed, corresponding to interface-normal compression under zero lateral strain. Although the nominal pressure is identical, the cell deformation differs between the two loading modes. Under uniaxial loading, the z-axis cell length decreases more rapidly than under isostatic loading, showing that compression is concentrated along the interface-normal direction (Figs. 4a,b).

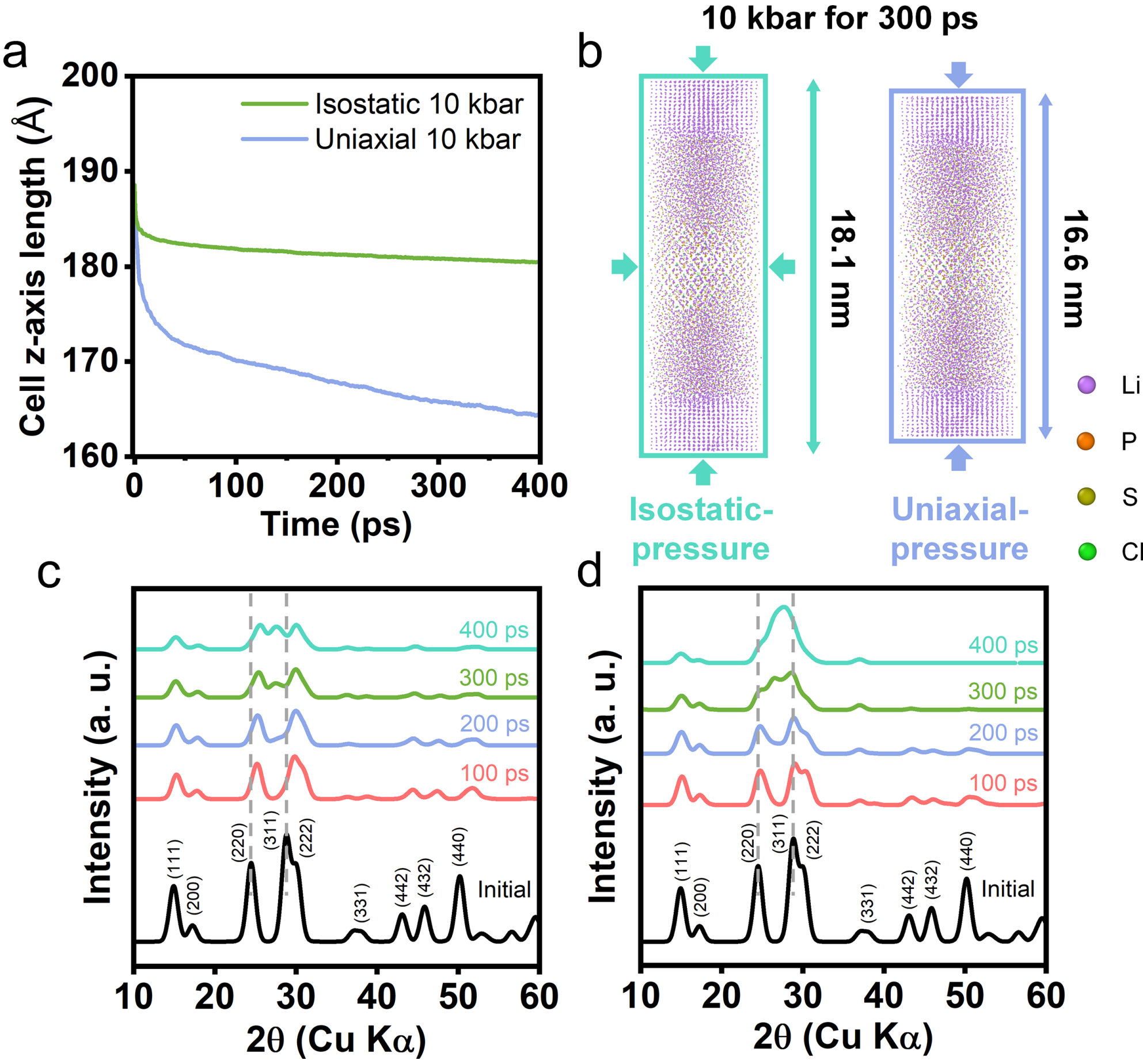

**Fig. 4 I Isostatic and uniaxial pressure effects on LiIILPSC interfacial reactions.** (a) Time evolution of the z-axis cell length under 10 kbar isostatic and 10 kbar uniaxial pressure in DP-QEq ConstQ MD. (b) Representative structures after 300 ps under the two pressure-loading modes. (c,d) Time-resolved XRD patterns from 0 to 400 ps under (c) 10 kbar isostatic pressure and (d) 10 kbar uniaxial pressure. The indexed reflections in the initial pattern correspond to the crystallographic planes of pristine LPSC.

The different cell response is accompanied by different interfacial chemistry. The $PS_4$ tetrahedra destruction ratio increases more rapidly under uniaxial compression than under isostatic compression (Supplementary Fig. 37), indicating accelerated interfacial decomposition. Time-dependent XRD results show that the development of $Li_2S$-like crystalline ordering proceeds faster under uniaxial compression, with a clearer $Li_2S$-like reflection appearing by 400 ps (Figs. 4c,d). Long-time DP simulations show the same trend (Supplementary Figs. 15b and 38b). At 10 kbar, extensive crystallization remains suppressed relative to the 1 kbar cases, but uniaxial loading still promotes faster interfacial degradation than isostatic loading (Supplementary Fig. 38). Therefore, pressure magnitude alone is insufficient to describe Li||LPSC interfacial evolution; the loading geometry also affects electrolyte deformation and interfacial reaction.

A notable difference also appears in the LPSC lattice response. Under isostatic pressure, the LPSC bulk peaks shift more clearly to higher diffraction angles, indicating more uniform lattice contraction throughout the electrolyte region (Figs. 4c,d). Under uniaxial pressure, the same nominal pressure produces negligible contraction of the remaining LPSC lattice but stronger interfacial decomposition and $Li_2S$-like peak growth. These results indicate that isostatic and uniaxial compression distribute mechanical deformation differently between the electrolyte bulk and the reacting interface. Such loading method-dependent lattice contraction can influence bulk Li-ion transport, while the accompanying difference in interfacial decomposition can alter SEI growth and interfacial resistance. Although real cells contain additional heterogeneities, this comparison isolates how pressure anisotropy can couple lattice deformation with interfacial chemistry.

**Defect location determines pressure-driven interfacial evolution**

Void formation and dead Li are major degradation modes in ASSLMBs because they reduce contact area, increase current constriction, promote local overpotentials, and can allow electronically isolated Li to continue reacting chemically with sulfide electrolytes[15, 36, 37, 38]. Recent simulations and experiments have suggested that Li clusters can form within chemically modified interphases and that voids or pores in the interphase can participate in dendrite-related degradation[24, 39]. We therefore used DP-QEq ConstQ simulations to examine how pressure affects voids and dead-Li clusters at different locations in the Li||LPSC system. Voids were placed in the LPSC bulk side, on the Li-metal side of the interface, and on the LPSC side of the interface (Figs. 5a, Supplementary Figs. 39–41). Embedded Li clusters were generated by inserting Li metal into voids in the LPSC bulk and at the LPSC-side interface and were used as structural analogues of dead-Li domains (Extended Data Fig. 1a and Supplementary Fig. 42a).

Void closure is strongly location-dependent. For a void in the LPSC bulk, 1 bar and 1 kbar lead to slow volume reduction, whereas 10 kbar closes the void within 200 ps (Supplementary Figs. 39 and 40). For a void on the Li-metal side of the interface, the void shrinks faster at higher pressure and closes more rapidly than a void in LPSC bulk (Supplementary Fig. 41). In the 1 kbar simulation, the Li-side void disappears on the tens-of-picoseconds time scale, while the LPSC-side bulk void remains substantially open over hundreds of picoseconds. This behavior is consistent with the lower hardness and greater deformability of Li metal relative to the sulfide electrolyte.

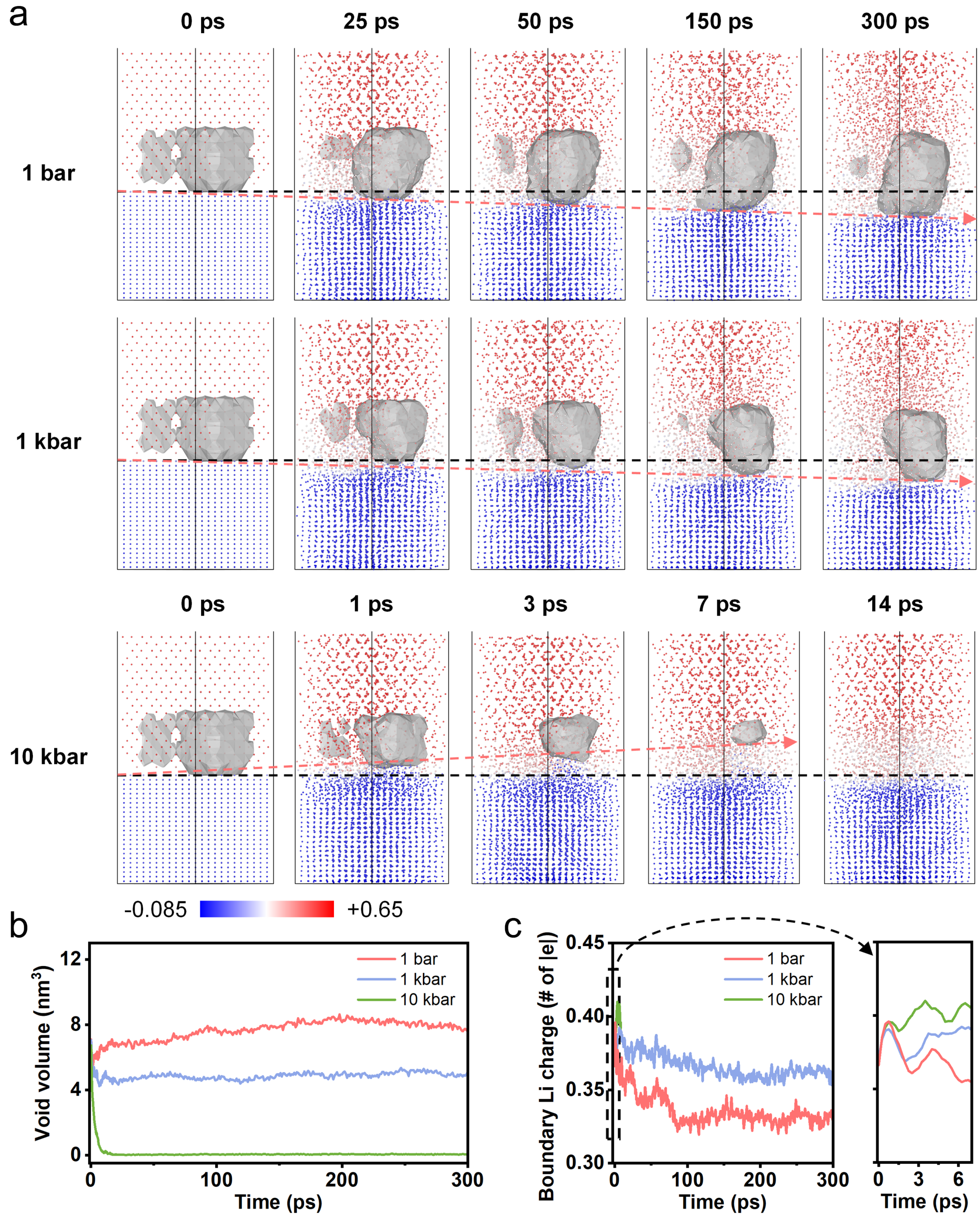

**Fig. 5 | Pressure-regulated evolution of an interfacial void at the Li|LPSC interface.** (a) Time-resolved DP-QEq ConstQ snapshots of an LPSC-side interfacial void under 1 bar, 1 kbar, and 10 kbar. The gray surface denotes the void, the color scale denotes local Li charge, and the dashed lines indicate the initial and evolving interfacial boundaries. (b) Void volume as a function of time under different pressures. (c) Mean charge of Li atoms at the void boundary as a function of time, with an early-time inset.

The LPSC-side interfacial void behaves differently from both bulk-LPSC and Li-side voids. At 10 kbar it closes rapidly, whereas at 1 kbar it does not close effectively and at 1 bar it even expands (Figs. 5a,b). This behavior occurs because the void locally blocks direct Li|LPSC contact. As the surrounding interface reacts and expands, Li beneath the void becomes partially isolated from the

electrolyte, while reaction proceeds around the void edge. The evolving interphase boundary can therefore move around and undercut the void, increasing the apparent void region instead of closing it. Within the pressures examined, effective closure of the LPSC-side interfacial void emerges only at 10 kbar; the same pressure that closes a Li-side void may thus be insufficient for a void embedded in the reacting LPSC-side boundary.

The mean charge of Li atoms at the void boundary provides a useful descriptor of the void-closure pathway (Figs. 5c and Supplementary Fig. 41c). Li in metallic environments has a charge close to neutral, whereas Li in the LPSC or reduced interphase is more positively charged because it is coordinated by anions. For an LPSC-side interfacial void, a decrease in boundary Li charge indicates that the void boundary is moving toward a more Li-metal-like environment; high pressure suppresses this motion and maintains boundary Li more positively charged. For a Li-side interfacial void, the opposite trend is observed: under stronger pressure, the void rapidly moves toward the LPSC/interphase side, increasing the boundary Li charge. Thus, pressure regulates not only the kinetics of void closure, but also the direction and chemical pathway of void evolution.

Dead-Li clusters respond to pressure through reactive spreading. The clusters gradually expand instead of retaining their initial compact metallic shape, accompanied by reaction with the surrounding electrolyte[37, 38] (Extended Data Fig. 1a and Supplementary Fig. 42a). Increasing pressure suppresses this spreading, as shown by reduced growth of the cluster volume (Extended Data Fig. 1b and Supplementary Fig. 42b) and a smaller increase in the radius of gyration (Extended Data Fig. 1c and Supplementary Fig. 42c). Analysis of the average Li charge within the dead-Li cluster further shows that high pressure better preserves a more metallic character of the dead-Li cluster, whereas lower pressure allows more extensive electron transfer to the surrounding sulfide framework (Extended Data Fig. 1d and Supplementary Fig. 42d). The pressure-induced suppression of dead-Li spreading is consistent with reduced interfacial reaction of the isolated Li cluster, although it does not imply recovery of electronic connectivity. These results show that pressure can limit the chemical expansion of dead Li, which is accompanied by electron transfer from metallic Li to the surrounding LPSC electrolyte.

## Discussion

Stack pressure at reactive Li||LPSC interfaces influences chemical evolution as well as physical contact. Within the simulated range, 1 kbar accelerates $PS_4$ decomposition and $Li_2S$-like ordering relative to 1 bar, whereas 10–100 kbar retain more of the LPSC framework and suppress long-range crystallization over the simulated timescale. Charge-resolved trajectories connect the early interphase to sulfur-centered, lithium-rich environments and show that pressure also changes local

P/Cl coordination and lithium mobility. These results identify a nonmonotonic competition between reaction, structural rearrangement and transport under compression.

The pressure response further depends on how mechanical loading is applied and where defects are located. Uniaxial compression concentrates deformation normal to the interface and accelerates decomposition compared with isostatic loading at the same nominal pressure. Pressure also promotes void closure but suppresses the reactive spreading of dead lithium, with the pathway determined by whether the defect lies in lithium metal, LPSC or the reacting boundary. Pressure magnitude alone is therefore insufficient to describe the evolution of a heterogeneous solid-state interface.

The simulations do not define an optimum operating stack pressure. They use an idealized Li||LPSC interface, accelerated dynamics at 400 K and pressure regimes ranging from high assembly and densification conditions to an extreme 100 kbar mechanistic limit; quantitative thresholds will depend on temperature, timescale, microstructure and chemistry. Nevertheless, the combined results show that pressure should be treated as a mechanochemical variable coupling interphase chemistry, transport and morphology, and the nonmonotonic response cautions against extrapolating pressure effects from a single loading condition. Methodologically, incorporating the long-range electrostatic contribution into the virial extends charge-resolved machine-learning dynamics from fixed-volume to pressure-controlled ensembles, offering a transferable route to the mechanochemistry of reactive, electrified solid–solid interfaces; the resulting pressure-dependent trends will require system-specific validation.

## Methods

### Density functional theory and ab initio molecular dynamics

All DFT calculations were performed using the Vienna Ab initio Simulation Package (VASP)[40] with the projector-augmented-wave[41, 42] method and the Perdew-Burke-Ernzerhof[43] generalized-gradient approximation. A plane-wave energy cutoff of 500 eV was used. Electronic self-consistency was converged to $1.0 \times 10^{-6}$ eV, and geometry optimizations were performed until the residual forces were below 0.02 eV $Å^{-1}$. Gaussian smearing with a width of 0.05 eV was used. The Brillouin zone was sampled using a 2 × 2 × 1 Monkhorst-Pack k-point mesh for the interfacial cells, while denser k-point meshes were used for the bulk cells. Spin polarization was included in the DFT calculations. The optimized structures and single-point DFT energies, forces, and virials were used as reference labels for machine-learning interatomic-potential training and validation.

AIMD simulations were performed to generate initial training configurations for Li||LPSC interfaces, modified Li||LPSC interfaces ($LPSC_{vac}$ and $LPSC_{s}$), LPSC bulk, and Li bulk. AIMD datasets were generated through sequential DFT relaxation, 1 ps NVT equilibration with a 1 fs time step, and 20 ps NPT production simulations with a 2 fs time step. The NPT production trajectories were propagated using a Langevin thermostat combined with Parrinello–Rahman cell dynamics. For each system, seven AIMD production trajectories were collected: four non-external-pressure trajectories at 400, 600, 800, and 1000 K, and three additional 400 K trajectories under external pressures of 1, 5, and 10 kbar. For externally pressurized AIMD trajectories, the reported energies included a positive pressure-volume contribution. This contribution was extracted for each configuration and subtracted before constructing the MLIP training labels. Each 20 ps NPT trajectory produced 10,000 snapshots. To reduce redundancy while preserving structural diversity, every third snapshot was first selected from each trajectory, yielding 3,334 snapshots per trajectory and 23,338 snapshots per system. SOAP-based farthest-point sampling[44, 45] (FPS) was then applied to each pooled system-specific dataset to select structurally diverse subsets for training. The final numbers of SOAP-FPS-selected AIMD configurations are summarized in Supplementary Table 1. SOAP descriptors were calculated with periodic boundary conditions using a 5.5 Å radial cutoff, six radial basis functions to expand the radial environment, and spherical-harmonic angular components up to angular momentum $l = 4$.

### X-ray diffraction simulations

Simulated XRD patterns were calculated from MD snapshots using the pymatgen[46] XRDCalculator with Cu Kα radiation. For each structure, diffraction peaks were generated within the selected 2θ range and mapped onto a uniform 2θ grid. To mimic finite-temperature and finite-size broadening, each discrete peak was convoluted with a Gaussian function. A Gaussian width of $\sigma = 0.3°$ was used. For time-resolved or region-resolved analyses, XRD patterns were calculated from the corresponding sampled structures or structural subsets. When multiple MD frames were used for one time point or condition, the broadened intensities were averaged over all selected frames before plotting.

### Interphase-region assignment

For region-resolved analyses, the interface-normal direction was taken as the z axis and the cell was divided into 80 bins along z. In each bin, Li, P, S, and Cl atoms were counted. Because P, S, and Cl originate from the LPSC framework, the local LPSC-framework atom fraction, $f_{framework}(\mathrm{z})$ was used as a composition marker:

$$f_{framework}(\mathrm{z}) = \frac{N_P(z) + N_S(z) + N_{Cl}(z)}{N_{Li}(z) + N_P(z) + N_S(z) + N_{Cl}(z)}$$

where $N_i(z)$ is the number of element $i$ in the corresponding z bin. Bins with $0.05 \leq f_{framework}(\mathrm{z}) \leq 0.50$ were assigned as compositionally mixed interphase bins, whereas Li-rich and LPSC-framework-rich bins outside the interphase were assigned to the Li metal bulk and LPSC bulk regions, respectively. Consecutive mixed bins were grouped into continuous interphase segments. Because the periodic Li||LPSC slab contains two Li||LPSC contacts, the lowest-z and highest-z mixed segments were used as the lower and upper SEI/interphase slabs, respectively. This procedure was repeated independently for every frame so that changes in cell length and interphase growth under pressure were tracked dynamically.

**Deep Potential and DP-GEN models**

The DP model was trained using DeePMD-kit v2.2.10[26]. A smooth edition two-body embedding descriptor (se_e2_a) was used with a cutoff radius of 5.5 Å and a smoothing cutoff of 0.5 Å. The maximum neighbor selections for Li, P, S, and Cl were 70, 12, 35, and 12, respectively. The embedding network contained three hidden layers with 25, 50, and 100 neurons, and the fitting network contained three hidden layers with 240 neurons each. The model was trained for 2,000,000 steps using an exponentially decaying learning rate from $1.0 \times 10^{-3}$ to $1.0 \times 10^{-7}$. The loss function included energy, force, and virial terms. The force prefactor was decreased from 1000 to 1.2 during training, while the energy and virial prefactors were increased from 0.02 to 1.0 and 1.2, respectively. These settings are optimized for training an accurate force-dominated model while retaining energy and stress consistency.

Active learning was performed using the DP-GEN[27] concurrent-learning workflow, consisting of iterative training, exploration, model-deviation selection, and DFT labeling. In DP-GEN, an ensemble of independently trained models is used to estimate the uncertainty of the potential during MD exploration. Configurations for which the model ensemble gives similar force and virial predictions are considered well represented by the current training set, whereas configurations with large model deviations indicate under-sampled regions of configuration space. These uncertain configurations are selected for DFT labeling and added back to the training set in the next iteration.

Four independently initialized DP models were trained in each iteration. Candidate configurations were generated by MD exploration, and model deviation in forces and virials was used to identify configurations for DFT labeling. For Li and LPSC bulk active learning, force-deviation trust levels of approximately 0.03–0.15 eV $Å^{-1}$ for Li and 0.04–0.15 eV $Å^{-1}$ for LPSC were used, with virial-deviation trust windows of 0.05–0.15 and 0.06–0.15 eV $atom^{-1}$, respectively. For Li||LPSC interfacial active learning, a wider force-deviation window of 0.15–0.35 eV $Å^{-1}$ and a virial-deviation window of 0.008–0.015 eV $atom^{-1}$ were used to capture the broader chemical diversity of the reacting interface. MD exploration was carried out progressively at 400 K by starting from 1 kbar pressure conditions and gradually increasing the external pressure up to 150 kbar. This staged exploration allowed the active-learning loop to first sample relatively certain configurations near the initial training domain and then expand toward lower-certainty, high-pressure configurations. Configurations with model deviations within the labeling window were selected for DFT calculations and added to the training set, enabling reliable coverage of increasingly compressed bulk and interfacial structures.

**DP-QEq model**

The DP-QEq[24] model was used to describe Li||LPSC interfacial dynamics in which local bond rearrangement and long-range charge redistribution occur simultaneously. A conventional DP model learns the local many-body potential-energy surface from atomic environments, but it does not explicitly assign environment-dependent atomic charges. DP-QEq extends this framework by combining a short-range DP model with an on-the-fly charge-equilibration model. At each MD step, $DP_{short}$ describes the local residual interaction, while the QEq solver determines atomic charges that minimize the electrostatic energy under the selected electrochemical boundary condition. The optimized charges provide long-range electrostatic energy, forces, virial/stress.

The total DP-QEq energy is written as the sum of a short-range DP contribution and a long-range QEq contribution,

$$E_{Total}(R, q) = E_{Short}(R) + E_{QEq}(R, q)$$

where R denotes the atomic coordinates and q denotes the atomic charges. The QEq energy contains electrostatic interactions, element-dependent electronegativity terms, and atomic hardness terms,

$$E_{QEq} = E_{Coulomb} + \sum_{i=1}^{N} \left( \chi_i^0 Q_i + \frac{1}{2} J_i Q_i^2 \right)$$

where $Q_i$ is the charge, $\chi_i$ is the electronegativity, and $J_i$ is the hardness of atom $i$. $E_{Coulomb}$ was evaluated using particle-mesh Ewald[47] electrostatics with Gaussian self-correction and dipole-correction terms.

For ConstQ simulations, the charges were optimized by minimizing $E_{QEq}$ under a fixed total-charge constraint.

$$Q_{cell} = \sum_{i=1}^{N} Q_i$$

This condition gives equalized electrochemical potential across the simulation cell while conserving the prescribed total charge. The charge variables were optimized using an LBFGS algorithm solver with a projected-gradient tolerance of $1.0 \times 10^{-2}$. Element-specific QEq parameters were taken from the calibrated input[48]: the electronegativity parameters for Li, P, S, and Cl were -3.0, +1.8, +6.5745, and +10, respectively. The hardness parameters for Li, P, S, and Cl were 10.0241, 7.0946, 9.0000, and 6.0403.

For ConstP simulations, the QEq minimization was modified by applying electrode-dependent electronegativity shifts to Li atoms belonging to the metallic electrodes,

$$\chi_i = \chi_i^0 + \Delta\chi_{electrode}$$

where, $\Delta\chi_{electrode}$ denotes the electronegativity shift assigned to the electrode containing atom $i$. Opposite electronegativity shifts impose an electrode bias and allow charge to redistribute between the two Li electrodes according to the modified electrochemical boundary condition.

After charge optimization, the QEq contribution was combined with $DP_{short}$ at the level of energy, forces, and virial,

$$E_{tot} = E_{short} + E_{QEq}$$

$$F_{tot} = F_{short} + F_{QEq}$$

$$V_{tot} = V_{short} + V_{QEq}$$

Including the QEq virial is important for NPT simulations because the long-range electrostatic term contributes directly to the stress response of the charged interface.

**Pressure-coupled DP-QEq implementation**

The DP-QEq implementation used in this work was adapted from the previously reported ConstQ/ConstP framework and extended for pressure-controlled Li||LPSC interfacial simulations. First, the chemical type map and QEq parameter tables were replaced with Li, P, S, and Cl parameters, and the neighbor-list and pair-buffer settings were enlarged for LPSC bulk and Li||LPSC interfacial cells. The PME grid and dipole-correction treatment were also adjusted for the anisotropic Li||LPSC simulation cells. The DeepMD interface was also updated to the DeePMD-kit v2 DeepPot.eval interface. Second, the QEq contribution was extended to include virial and stress. After charge optimization, the $DP_{short}$ and QEq contributions were combined at the level of energy, forces, and virial. The QEq virial was obtained from the cell derivative of the QEq energy, added to the $DP_{short}$ virial, converted to the ASE stress convention, and returned together with the total energy and forces. This modification is required for NPT simulations because the long-range electrostatic contribution directly affects the stress response of the charged Li||LPSC interface. Third, the MD driver was modified to perform pressure-controlled DP-QEq simulations using the ASE NPTBerendsen barostat[49]. The updated implementation reads the target pressure and thermostat/barostat coupling parameters from the input file and couples the DP-QEq energy, forces, and stress to NPT cell evolution. This modification enables ConstQ DP-QEq simulations of Li||LPSC interfaces under externally applied pressure. Early-stage pressure-controlled DP-QEq ConstQ simulations were performed at 400 K with a 1 fs time step.

### Production-trajectory DFT validation

Production validation was conducted using 200 ps DP-driven NPT molecular dynamics simulations of the Li||LPSC interface cell (Supplementary Fig. 1a) at 0 bar, 1 bar, 1 kbar, 5 kbar, 10 kbar, and 100 kbar. All simulations were performed at 400 K to facilitate interfacial evolution within the simulated timescale and enable pressure-dependent comparisons under a common thermal condition. From each pressure-resolved trajectory, 200 snapshots were uniformly sampled and independently labeled by DFT single-point calculations. The resulting energy, force, and virial labels were used to evaluate both the DP and DP-QEq models against DFT on the same configurations at each pressure. The corresponding pressure-resolved RMSEs are summarized in Supplementary Tables 4 and 5 for the DP and DP-QEq models, respectively. Figs. 1b and 1c combine the validation data from all pressure conditions for the DP–DFT and DP-QEq–DFT comparisons, respectively.

### Isostatic and uniaxial pressure simulations

To compare pressure-loading modes, 10 kbar isostatic and 10 kbar uniaxial DP-QEq ConstQ NPT simulations were performed using the same Li||LPSC interfacial structure. Both simulations used 400 K, a 1 fs time step, Berendsen thermostat and barostat coupling[50], and a compressibility of 5.0 × 10-6 $bar^{-1}$.

The loading modes differed in the barostat degrees of freedom. In the isostatic case, the standard ASE NPTBerendsen barostat[49] was used with the scalar pressure set to 10 kbar, so the simulation cell was pressure-coupled as a whole and the in-plane and interface-normal lattice vectors were allowed to respond together to the external pressure. In the uniaxial case, the barostat was changed to ASE Inhomogeneous_NPTBerendsen with the same nominal pressure, temperature, coupling times, and compressibility, but with a cell-scaling mask of (0, 0, 1). This mask restricts pressure-driven cell scaling to the z direction, corresponding to the interface-normal direction, while keeping the x and y cell dimensions fixed. Because the lateral cell dimensions are fixed, this protocol represents interface-normal compression under zero lateral strain rather than a uniaxial-stress condition with free lateral relaxation. The comparison therefore separates hydrostatic densification of the full simulation cell from directional compression normal to the Li||LPSC interface.

### Void and dead-lithium models

Void simulations were performed by introducing voids at three locations: the LPSC-side interfacial region, the Li-metal-side interfacial region, and the LPSC bulk side. For voids formed in the LPSC region, atoms were removed as stoichiometric $Li_6PS_5Cl$ units so that the composition of the remaining LPSC framework was not changed. The removed region was selected to generate a void approximately 20 × 20 × 20 $Å^3$ in size, and the void position was chosen so that the distance between periodic images of the void was larger than 20 Å under periodic boundary conditions. For voids formed in the Li-metal region, Li atoms were removed from a similarly sized region while preserving the surrounding Li-metal lattice.

Dead-Li structures were generated from the preformed void geometries by filling the void region with Li metal. Two dead-Li locations were considered: a Li cluster embedded in the LPSC bulk and a Li cluster located at the LPSC-side interface. The inserted Li cluster was placed densely within the void while avoiding direct overlap with the pre-existing atoms in the simulation cell. The inserted Li atoms retained a Li-metal-like local spacing.

DP-QEq ConstQ NPT simulations were then performed under 1 bar, 1 kbar, and 10 kbar for the void systems. Dead-Li simulations were performed under 1 bar, 10 kbar, and 100 kbar for both the LPSC-bulk and LPSC-side interfacial dead-Li models. Void volumes were calculated using Zeo++[51] by identifying the connected empty volume associated with the initial void under periodic boundary conditions. Void morphology, void volume, and boundary Li charge were monitored as functions of time. For dead-Li simulations, pressure-dependent changes in cluster volume, radius of gyration ($R_g$), local environment, and Li charge were analyzed to assess how pressure suppresses spatial spreading and changes the electrochemical character of dead Li in different LPSC-side locations. Atomic configurations and MD trajectories were visualized using OVITO[52].

### Data availability

The training datasets and initial Li|LPSC interfacial structure are available on Zenodo at https://doi.org/10.5281/zenodo.22028882.

### Code availability

The simulation code used in this study is available in the GitHub repository at https://github.com/snu-micc/pressure-dpqeq.

## Acknowledgements

This work was supported by the National Research Foundation of Korea (NRF) (RS-2024-00429941)

## Author contributions

K.J. conceptualized the project and wrote the original draft. K.J. and J.C. performed the analyses. Y.J. supervised the project. All authors discussed the results and reviewed and edited the manuscript.

## Competing interests

The authors declare no competing interests.

# Extended Data

**Extended Data Fig. 1 | Pressure-dependent behavior of a dead-Li cluster embedded in the LPSC bulk region.** (a) Time-resolved snapshots under 1 bar, 10 kbar, and 100 kbar. The gray surface represents the selected Li cluster, and the color scale indicates Li charge. (b) Li cluster volume as a function of time. (c) Radius of gyration ($R_g$) of the Li cluster. (d) Mean Li cluster charge at early times.